\documentclass[12 pt]{article}

\usepackage{feynmf}
\usepackage{rotating}
\newcommand{\approxge}{\stackrel{>}{\scriptstyle\sim}}
\newcommand{\approxle}{\stackrel{<}{\scriptstyle\sim}}

\begin{document}\setlength{\unitlength}{1mm}

\begin{flushright}
\begin{tabular}{l}
AMES-HET-97-02 \\
March 1997
\end{tabular}
\end{flushright}
\begin{center}
\vspace{12pt}
\LARGE{Direct Top Quark Production at Hadron Colliders as a Probe of New
Physics} \\
\vspace{12pt}
\Large{M. Hosch, K. Whisnant, and B.-L. Young} \\
Department of Physics and Astronomy \\
Iowa State University \\
Ames, Iowa 50011 USA \\
\vspace{12pt}
\end{center}

\begin{abstract}
We examine the effect of an anomalous flavor changing chromomagnetic
moment which allows direct top quark production (two partons combining
into an unaccompanied single top quark in the s-channel) at hadron
colliders.  We consider both t-c-g and t-u-g couplings.  We find that
the anomalous charm quark coupling parameter $\kappa_c / \Lambda$ can be
measured down to $.06\rm{~TeV}^{-1}$($.009 \rm{~TeV}^{-1}$) at the
Tevatron with the Main Injector upgrade(LHC).  The anomalous up quark
coupling parameter $\kappa_u / \Lambda$ can be measured to $.02
\rm{~TeV}^{-1}$($.003 \rm{~TeV}^{-1}$) at the Tevatron(LHC).
\end{abstract}

\section*{Introduction}
With the discovery of the top quark \cite{cdftop,d0top}, the long
anticipated completion of the fermion sector of the standard model has
been achieved.  Its unexpected large mass in comparison with the other
known fermions suggests that the top quark may play a unique role in
probing new physics, and has prompted both theorists and experimenters
alike to search for anomalous couplings involving the top quark.  On the
experimental side, the CDF \cite{Incandela,LeCompte} and D0
\cite{Heinson} collaborations have begun to explore the physics of top
quark rare decays \cite{Incandela}.  On the theoretical side, a
systematic examination of anomalous top quark interactions, in a model
independent way, has been actively undertaken\cite{Han1,Rizzo}.

One possible set of anomalous interactions for the top quark is given by
the flavor-changing chromo-magnetic operators:
\begin{equation}
\frac{\kappa_u}{\Lambda} g_s \overline{u} \sigma^{\mu \nu}
\frac{\lambda^a}{2} t G^a_{\mu \nu}~~+~~h.c.~,
\end{equation}
and
\begin{equation}
\frac{\kappa_c}{\Lambda} g_s \overline{c} \sigma^{\mu \nu}
\frac{\lambda^a}{2} t G^a_{\mu \nu}~~+~~h.c.~,
\end{equation}
where $\Lambda$ is the new physics scale, $\kappa_c$ and $\kappa_u$
define the strengths of the couplings, and $G^a_{\mu \nu}$ is the gauge
field tensor of the gluon.  The investigation of these couplings is well
motivated.  Although these operators can be induced in the standard
model through higher order loops, their effects are too small to be
observable\cite{Grzad}.  Therefore, any observed signal indicating these
types of couplings is 
direct evidence for physics beyond the standard model.

It has been argued that the couplings in Eqs. (1) and (2) may be
significant in many extensions to the standard model, such as 
supersymmetry (SUSY) or other models with multiple Higgs doublets
\cite{Grzad,Couture,Lopez,Cheng}, models with new dynamical interactions of the
top quark\cite{Hill}, and models where the top quark has a
composite\cite{Georgi} or soliton\cite{Zhang} structure.  In particular,
Ref. \cite{Lopez} suggests that the supersymmetric contributions to a
t-c-g vertex may be large enough to measure at a future hadron collider.

T. Han et. al.\cite{Han2} have placed a limit on the top-charm-gluon
coupling strength, $\kappa_c$, by examining the decay of the top quark
into a charm quark and a gluon.  They find an upper limit on $\kappa_c /
\Lambda$ of $.43(.65) \rm{~TeV}^{-1}$ with(without) b-tagging for $200
\rm{~pb}^{-1}$ of data at the Tevatron.  If the c and u jets are not
distinguished, their result applies equally well to $\kappa_u /
\Lambda$, if one uses the up quark coupling alone, or to the sum, added
in quadrature, when both are considered. 

In this paper, we will examine these operators in a model independent
way using direct top quark
production at the Fermilab Tevatron and at the CERN LHC.  In this
scenario, a charm (or up) quark and a gluon from the colliding
hadrons combine immediately to form an s-channel top quark, which then
decays.  The production of a single, unaccompanied top or anti-top quark
is very small in the standard model.  For simplicity, we consider
couplings to the up quark and to the charm quark independently.  We will
take as our signal only the case where the top quark decays to a b quark
and a W boson.  While the $t
\to c g \rm{~(or~}u g\rm{)}$ decay will occur in the presence of the
anomalous couplings given in Eqs. (1) and (2), it is smaller than the $t
\rightarrow b W$ decay for $\kappa / \Lambda \approxle .75
\rm{~TeV}^{-1}$, and will have a negligible branching ratio for $\kappa
/ \Lambda \approxle .2 
\rm{~TeV}^{-1}$.  Given the existing upper bound of the anomalous
coupling mentioned earlier \cite{Han2}, $t \to b W$ will be the dominant
decay mode of the top quark.  Since the W boson decay into a charged
lepton (electron or muon) and its corresponding neutrino 
has an identifiable signature, we consider only the $t \to b W \to b l
\nu_l$ decay for our signal.  With the decays so chosen, we find that
the backgrounds are manageable, as will be discussed in detail later.

\section*{Direct Top Quark Production}

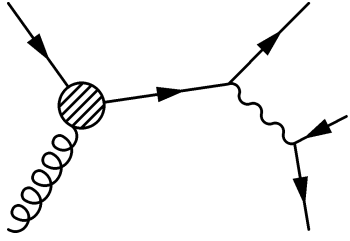
\begin{figure}
\begin{center}
\begin{fmffile}{kappa2}
\begin{fmfgraph*}(50,30)
\fmfleftn{i}{2}
\fmfrightn{o}{3}
\fmflabel{$g$}{i1}
\fmflabel{$c,u$}{i2}
\fmflabel{$b$}{o3}
\fmflabel{$l^+$}{o2}
\fmflabel{$\nu_l$}{o1}
\fmf{fermion}{i2,v1}
\fmf{gluon}{i1,v1}
\fmf{fermion,label=$t$}{v1,v2}
\fmf{fermion}{v2,o3}
\fmf{boson,label=$W^+$}{v2,v3}
\fmf{fermion}{o2,v3,o1}
\fmfblob{.12w}{v1}
\end{fmfgraph*}
\end{fmffile}
\end{center}
\caption{\label{feynman} Feynmann diagram for direct top quark
production and subsequent decay into $b l \nu_l$}
\end{figure}

We have calculated tree level cross sections for direct top quark
production, $p \overline{p} \rightarrow t \rightarrow b W^{+} \to b l^+
\nu_{l}$, using the flavor-changing chromomagnetic moments in Eqs. (1)
and (2) (see Fig.
\ref{feynman}).  The $l^+$ in this process is either a positron or an
anti-muon, and $\nu_l$ is its corresponding neutrino.  We also included
direct anti-top production in our calculation ($p \overline{p} \to
\overline{t} \to \overline{b} W^- \to \overline{b} l^-
\overline{\nu_l}$).  The parton cross section for direct top(or
anti-top) production is given by:
\begin{equation}
d \sigma = \frac{1}{4} \frac{1}{\left( 4 \pi \right)^5} \frac{\hat{s} -
M_{l,\nu_l}^2}{\hat{s}^2} |\overline{\mathcal{M}}|^2 d \Omega_b d
\Omega_l d M_{l,\nu_l}^2 ~,
\end{equation}
where the spin averaged squared matrix element is
\begin{equation}
|\overline{\mathcal{M}}|^2 = \frac{256 \pi^3 \alpha_2^2
\alpha_s}{3}  \frac{\kappa_{c(u)}^2}{\Lambda^2} 
\frac {\hat{s}
\left(p_b \cdot p_{\nu_l} \right) \left[ \hat{s}
\left(q_{c(u)} \cdot p_l \right) + m_t^2 \left(
q_g \cdot p_l \right) \right]} {\left( \left( \hat{s}
- m_t^2 \right)^2 + m_t^2 \Gamma_t^2 \right) \!\! \left( \left(
M_{l,\nu_l}^2 - M_W^2 \right)^2 + M_W^2 \Gamma_W^2 \right)}~,
\end{equation}
$p_{b,l,\nu_l}$ are the 4-momenta of the outgoing b quark,
lepton, and neutrino respectively, $q_{c(u),g}$ are the 4-momenta
of the incoming charm(up) quark and gluon,
$\Gamma_W$ is the decay width of the $W$ boson,
\begin{equation}
\Gamma_t = \Gamma_{t \to b W} \left[ 1 + \frac{128 M_W^2 \alpha_s}{3
\alpha_2 \left( 1 - \frac{M_W^2}{m_t^2}\right)^2 \left( 1 + 2
\frac{M_W^2}{m_t^2} \right) } \left( \frac{\kappa_{c(u)}}{\Lambda}
\right)^2 \right]
\end{equation}
is the decay width of the top quark, including the anomalous
contribution for $t \to c g$ (or $t \to u g$), $\Gamma_{t \to b W}$ is
the standard model top quark decay width to a $b$ quark and $W$ boson,
\begin{equation}
M_{l,\nu_l}^2 \equiv \left( p_l + p_{\nu_l} \right)^2
\end{equation}
is the invariant mass squared, not necessarily on shell, of the W boson,
and $\sqrt{\hat{s}}$ is the parton center of mass energy.

\begin{figure}
\begin{center}
%\begin{turn}{270}
%\includegraphics[scale=.4]{kappa.eps}
\includegraphics{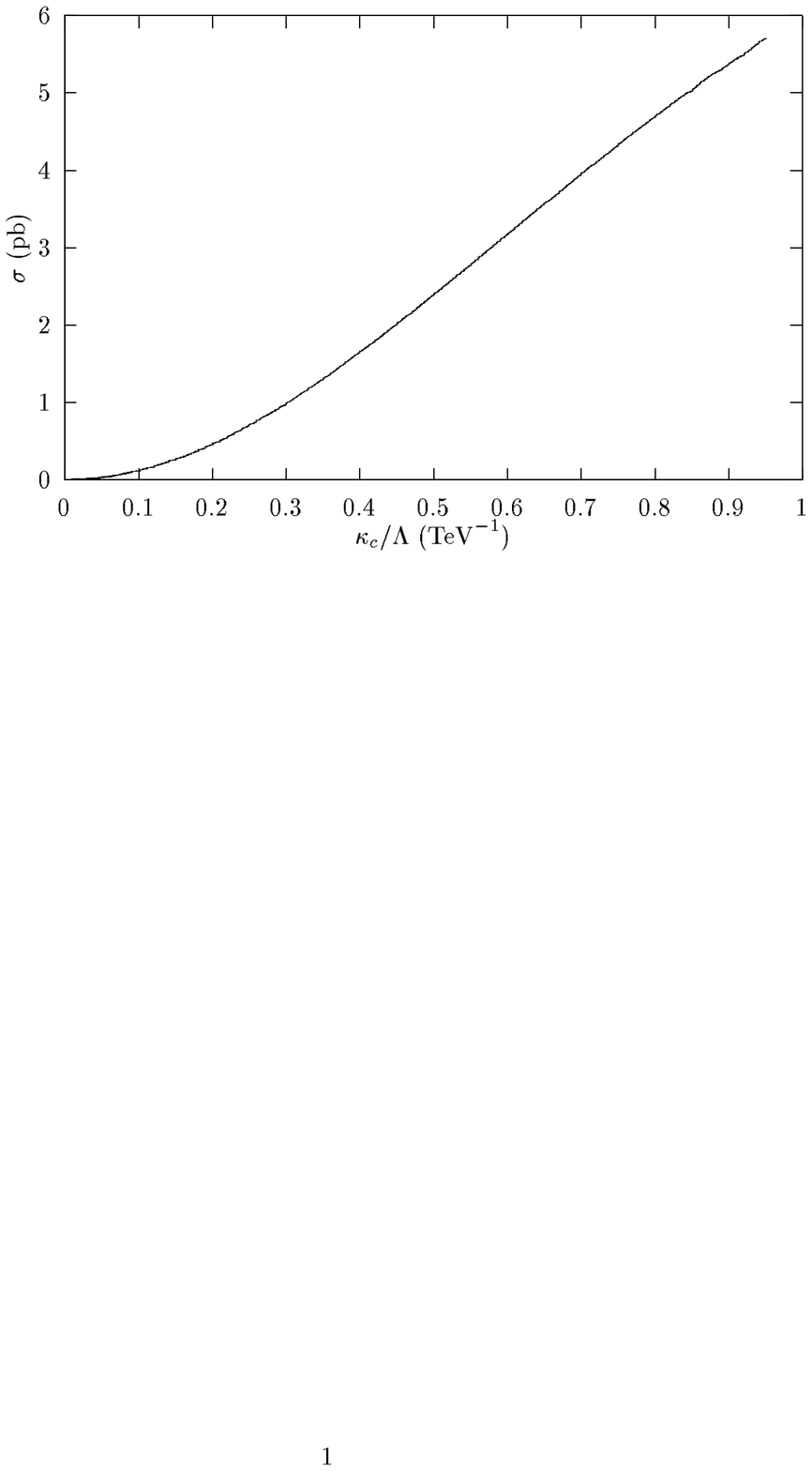}
%\end{turn}
\end{center}
\caption{\label{kappa} Direct top cross section vs. $\kappa_c /
\Lambda$ at Run 1 of the Tevatron.}
\end{figure}

As mentioned earlier, we considered only the case which has a charged
lepton (muon or 
electron) in the final state, to identify the $W$ boson.  Compared to
the hadronic decay mode of the $W$, the 
background for these processes is smaller and the signal is not as hard
to identify.  In order to examine the kinematics of the decay products,
we calculated the full three body phase space for the process, using the
Breit-Wigner propagators to broaden the top quark 
and W boson distributions.  Figure \ref{kappa} shows the
cross section at the Tevatron as a function of $\kappa_c / \Lambda$.  In 
the top quark decay width, we included an 
additional term arising from $t \rightarrow c g$, as shown in Eq. (5).
This term is proportional to $| \kappa_c / \Lambda |^2$ and 
contributes significantly to the top quark width only if $ \kappa_c /
\Lambda \approxge 0.2 \rm{TeV}^{-1}$.  
One can clearly see the effect of the additional channel for top quark
decay, which decreases the $t \to b W$ branching ratio and causes
a noticeable deviation from quadratic behavior for $\kappa_c / \Lambda
\approxge 0.2 \rm{~TeV}^{-1}$.  The corresponding cross section for the
t-u-g coupling has a similar shape, but is approximately an order of
magnitude larger.

We calculated the $p \overline{p}$ (for the Tevatron) and $p p$ (for the
LHC) cross sections for direct top production with the MRSA structure
functions \cite{MRS}.  We  
have also examined the effect of using the CTEQ3M \cite{CTEQ} structure
functions.  The difference between the two sets of structure functions
is small.  Several distributions were 
calculated, including the transverse momenta, the pseudorapidities, the
jet separation, from the lepton, and the reconstructed $\sqrt{\hat{s}}$.  

In order to reduce the $W + 1 \rm{~jet}$ background, we made a series of
cuts, which we will call the basic cuts,  on the kinematic
distributions.  They are:
\begin{eqnarray}
p_T  (b,l,\nu_l) &\ge& 25 \rm{~GeV} ~,\\ 
\eta_b &\le& 2.0 ~,\\
\eta_l &\le& 3.0 ~,\\
\Delta R &\ge& 0.4 ~,
\end{eqnarray}
where $\eta_{b,l}$ are the pseudorapidities, $\Delta R \equiv \sqrt{
\left(\eta_b - \eta_l \right)^2 + \left(\phi_b - \phi_l \right)^2}$ is the
separation between the b jet and the charged lepton in the detector, and
$\phi_{b,l}$ are the azimuthal angles.
We also assumed a Gaussian smearing of the energy of the final state
particles, given by:
\begin{eqnarray}
\Delta E / E & = & 30 \% / \sqrt{E} \oplus 1 \% \rm{,~for~leptons~,} \\
             & = & 80 \% / \sqrt{E} \oplus 5 \% \rm{,~for~hadrons~,} 
\end{eqnarray}
where $\oplus$ indicates that the energy dependent and independent
terms are added in quadrature.  

To enhance the signal relative to the background, we want to make cuts
on $\sqrt{\hat{s}}$, which should be sharply peaked at $m_t$ for the
signal.  To experimentally determine $\sqrt{\hat{s}}$, one must
reconstruct $p_t = p_b + p_l + p_{\nu_l}$.  The neutrino is not
observed, but its transverse momentum can be deduced from the
missing transverse momentum.  The longitudinal component of the 
neutrino momentum is determined by setting $M_{l,\nu_l} = M_W$ in Eq.
(6), and is given by:
\begin{equation}
p^{\nu_l}_L = \frac{\chi p^l_L \pm \sqrt{ \vec{p}_l^{\, 2} (\chi^2 -
p_{Tl}^2 p_{T\nu_l}^2) }}{p_{Tl}^2}~,
\end{equation}
where
\begin{equation}
\chi = \frac{M_W^2}{2} + \vec{p}^{\, l}_T \cdot \vec{p}^{\nu_l}_T ~,
\end{equation}
and $p_L$ and $p_T$ refer to the longitudinal and transverse momenta
respectively.   Note that there is a two fold ambiguity in this
determination.  We chose the solution which would best
reconstruct the mass of the top quark.  In some rare cases, the quantity
under the square root in Eq. (13) is negative.  When this happened, we
set this square root to zero, and used the corresponding result for the
neutrino longitudinal momentum.

\section*{Background Calculation}

\begin{figure}
\begin{center}
\begin{fmffile}{feynback2}
\begin{fmfgraph*}(50,30)
\fmfleftn{i}{2}
\fmfrightn{o}{2}
\fmflabel{$q$}{i1}
\fmflabel{$g$}{i2}
\fmflabel{$q'$}{o1}
\fmflabel{$W^+$}{o2}
\fmf{fermion}{i1,v1}
\fmf{fermion,label=$q$}{v1,v2}
\fmf{fermion}{v2,o1}
\fmf{gluon}{i2,v1}
\fmf{boson}{v2,o2}
\end{fmfgraph*}
\end{fmffile}
\end{center}
\caption{\label{backgd} Sample tree level Feynmann diagram for $W + 1
jet$ production}
\end{figure}
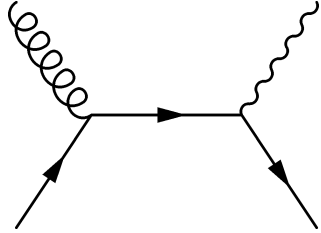

The main source of background to the direct top quark production is $p
\overline{p} \rightarrow W + 1 \rm{~jet}$.  A sample tree level Feynmann
diagram for this process is shown in
Fig. \ref{backgd}.  
Another background process is standard model single top
quark production when the associated jets are not observed.
Examining the data presented in Ref. \cite{Heinson2}, we conclude that
single top production is less than $1\%$ of the $W + 1
\rm{~jet}$ background when 
b-tagging is not used.  When b-tagging reduces the $W + 1
\rm{~jet}$ background by a factor of 100, the single top background may
be as large as $20\%$ of the total background.  However, since the
discovery limit on $\kappa / \Lambda$ scales as $B^{-\frac{1}{4}}$ where
B is the number of background events, a $20\%$ change in the background
affects the discovery limit by only $5\%$.  We therefore ignore this
background.

\begin{figure}
\begin{center}
\includegraphics[scale=.68]{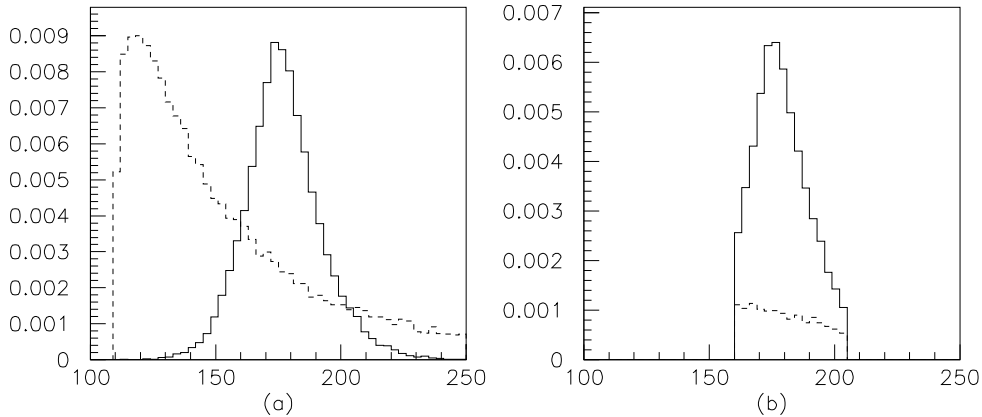}
\end{center}
\caption{\label{distrib1} $\sqrt{\hat{s}}$ distributions for the (a) basic
and (b) optimized cuts without b-tagging at the upgraded Tevatron.  The
solid line represents the direct top production($\kappa_c / \Lambda =
0.2 \rm{~TeV}^{-1}$).  The dashed line is one thousandth of the $W + 1
\rm{~jet}$ background.  The vertical axis is $d \sigma / d
\sqrt{\hat{s}}$ in pb/GeV, and the horizontal axis is $\sqrt{\hat{s}}$
in GeV.}
\end{figure}

\begin{figure}
\begin{center}
\includegraphics[scale=.68]{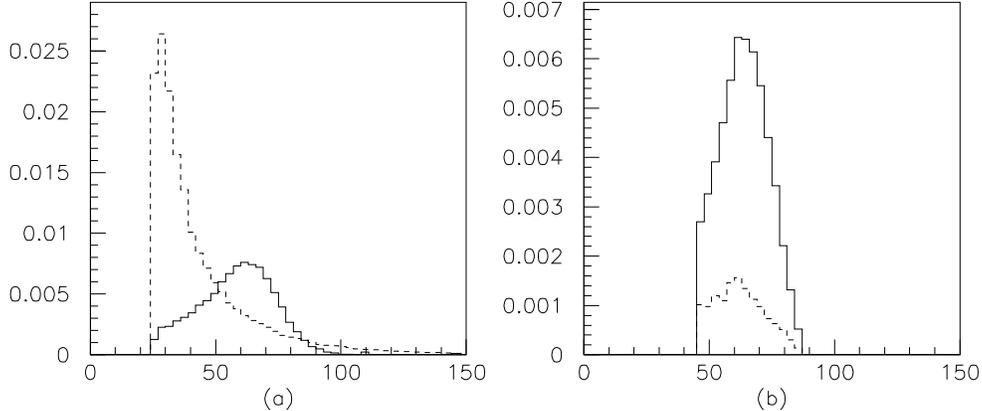}
\end{center}
\caption{\label{distrib2} $p_{Tb}$ distributions for the (a) basic
and (b) optimized cuts without b-tagging at the upgraded Tevatron.  The
solid line represents the direct top production($\kappa_c / \Lambda =
0.2 \rm{~TeV}^{-1}$).  The dashed line is one thousandth of the $W + 1
\rm{~jet}$ background.  The vertical axis is $d \sigma / d p_{Tb}$
in pb/GeV, and the horizontal axis is $p_{Tb}$ in GeV.}
\end{figure}

\begin{figure}
\begin{center}
\includegraphics[scale=.68]{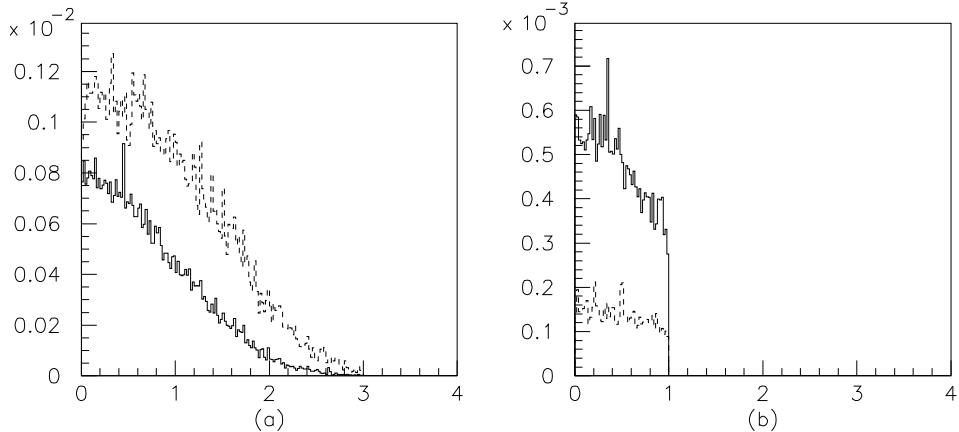}
\end{center}
\caption{\label{distrib3} $\eta_l$ distributions for the (a) basic
and (b) optimized cuts without b-tagging at the upgraded Tevatron.  The
solid line represents the direct top production($\kappa_c / \Lambda =
0.2 \rm{~TeV}^{-1}$).  The dashed line is one thousandth of the $W + 1
\rm{~jet}$ background.  The vertical axis is $d \sigma / d \eta_l$
in pb, and the horizontal axis is $\eta_l$.}
\end{figure}

We used the VECBOS monte-carlo \cite{vecbos} to calculate the cross
section for the  $W + 1 \rm{~jet}$ background.  We modified the program
to produce the same distributions that were calculated for the signal,
and applied the same basic cuts used in the signal calculation, Eqs.
(7-10).  To determine additional cuts which optimize the discovery limits on
$\kappa / \Lambda$, we examined the kinematic distributions in
$\sqrt{\hat{s}}$, $p_T$, $\eta$, and $\Delta R$.  We found that three
distributions, $\sqrt{\hat{s}}$, $p_{Tb}$, and $\eta_l$, were most
useful in isolating the signal from the background.  These are shown in
Figs. \ref{distrib1}, \ref{distrib2}, and \ref{distrib3}, with the charm
quark in the initial state and $\kappa_c / \Lambda = 0.2 \rm{~TeV}$
for the upgraded Tevatron.  The solid lines represent direct top
production, and the 
dashed lines represent the $W+1\rm{~jet}$ background divided by 1000.
The cuts were optimized for each of four cases: Run 1 at the Tevatron
with $p \overline{p}$ collisions at $\sqrt{s} = 1.8 \rm{~TeV}$ and $100
\rm{~pb}^{-1}$ of data per detector,  Run 2 with $\sqrt{s} = 2.0
\rm{~TeV}$ and $2 \rm{~fb}^{-1}$,  Run 3 with $2.0 \rm{~TeV}$ and $30
\rm{~fb}^{-1}$,  and the LHC with $p p$ collisions at $14 \rm{~TeV}$ and
$10 \rm{~fb}^{-1}$.  The optimized cuts are shown in Table \ref{cuts}. 
The corresponding distributions with the up quark in the initial state
are not shown; they have the same shape as for the charm quark, but are a
factor of ten larger in magnitude, due to the much larger size of the
valence up quark distribution in the initial state.

\begin{table}
\begin{center}
\begin{tabular}{|r|c|c|c|c|}
\hline
& $(p_{Tb})_{min}$ & $(\sqrt{\hat{s}})_{min}$ &
  $(\sqrt{\hat{s}})_{max}$ & $(\eta_l)_{max}$ \\ \hline \hline
1.8 TeV Tevatron & 35 GeV & 155 GeV & 205 GeV & 1.8 \\ \hline
2 TeV Tevatron   & 45 GeV & 160 GeV & 205 GeV & 1.0 \\ \hline
14 TeV LHC       & 35 GeV & 165 GeV & 195 GeV & 1.0 \\ \hline
\end{tabular}
\end{center}
\caption{\label{cuts} Optimized cuts for direct top quark
production}
\end{table}

To further reduce the background, we assumed that silicone vertex
tagging of the b jet would be available, with $36 \%$ efficiency at Run
1 of the Tevatron, and $60 \%$ at Runs 2 and 3, and at the LHC.  In
addition, we assumed that $1 \%$ of all non-b quark jets would be
mistagged as b quark jets. 

When b-tagging is present, if the jet produced is mistaken as a b jet,
it remains a part of the background.  The background can be reduced by a
factor of 100 if the $W + 1 \rm{~jet}$ sample does not include a
significant fraction of b quarks in the final state.  It is possible to
estimate the fraction of b quarks in the $W + 1 \rm{~jet}$ sample by
taking the ratio $|V_{cb}|^2 / |V_{ud}|^2$ and multiplying by the ratio
of the distribution fraction of charm quarks to up quarks in the
proton($\approx 0.1$). We estimate that the fraction of b quark jets in
the $W + 1 \rm{~jet}$ background is less than $.03\%$, much less than
the anticipated mistagging rate of $1 \%$.  We therefore ignore the
possibility of having b quarks in the $W + 1 \rm{~jet}$ sample.
Including b-tagging does not significantly affect the optimized cuts.

\section*{Results and Discussion}
\begin{table}
\begin{center}
\begin{tabular}{|l|r|r|r|r|r|}
\hline
&\multicolumn{1}{|c|}{$\sqrt{s}$}& \multicolumn{1}{c|}{Luminosity} &
\multicolumn{1}{c|}{background} &
\multicolumn{2}{c|}{signal needed (fb)} \\ \hline
& \multicolumn{1}{|c|}{TeV} & \multicolumn{1}{c|}{$\rm{fb}^{-1}$} &
\multicolumn{1}{c|}{fb} &
\multicolumn{1}{c|}{w/ b-tag} &\multicolumn{1}{c|}{w/o b-tag} \\
\hline
Run 1 & 1.8 & 0.1 & 19400 & 1370 & 190 \\ \hline
Run 2 & 2.0 & 2   & 13000 & 245  & 27  \\ \hline
Run 3 & 2.0 & 30  & 13000 & 63   & 6.4 \\ \hline
LHC   & 14  & 10  & 79000 & 267  & 27  \\ \hline
\end{tabular}
\end{center}
\caption{\label{cs} Signal needed for the discovery of anomalous t-c-g
and t-u-g couplings at the Tevatron and LHC at $95\%$ confidence level.
The background cross sections use the optimized cuts described in Table
\ref{cuts}.}
\end{table}

We can use the results of the signal and background calculations to
determine the minimum value of $\kappa_c / \Lambda$ or $\kappa_u /
\Lambda$ observable at hadron colliders.  Assuming Poisson statistics,
the number of signal events (S) required for discovery of a signal at
the $95\%$ confidence level is:
\begin{equation}
\frac{S}{\sqrt{S+B}} \ge 3 ~,
\end{equation}
where B is the number of background events obtained by multiplying the
background cross section by the luminosity and dividing by 100 if
b-tagging is present.  The luminosity, background cross section, and
signal cross section needed for discovery of anomalous flavor changing
couplings is given in Table \ref{cs}.  The discovery limits may then be
determined by comparing the signal calculation for a
given $\kappa / \Lambda$ to the signal needed, which can be obtained
from Table \ref{cs}.  These discovery limits are shown in Table
\ref{results}.

\begin{table}
\begin{center}
\begin{tabular}{|r|r|l|l|l|l|}
\hline
\multicolumn{2}{|c}{} & \multicolumn{3}{|c|}{Tevatron} &
\multicolumn{1}{c|}{LHC} \\
\cline{3-6}
\multicolumn{2}{|c|}{}& 1.8 TeV & \multicolumn{2}{|c|}{2 TeV} & 14 TeV
\\ \hline 
& b tagging? & $.1 fb^{-1}$ & $2 fb^{-1}$ & $30 fb^{-1}$ & $10 fb^{-1}$
\\ \hline \hline
charm 	& no 	& .38 	& .14   & .073  & .020   \\ \hline
	& yes	& .22	& .062	& .030	& .0084  \\ \hline \hline
u quark	& no	& .096	& .045	& .023	& .0081  \\ \hline
	& yes	& .058	& .019  & .0094 & .0033  \\ \hline
\end{tabular}
\end{center}

\caption{\label{results} Discovery limits on $\kappa_c / \Lambda$ and
$\kappa_u / \Lambda$ at the Tevatron and LHC.  The results are reported
in $\rm{TeV}^{-1}$}
\end{table}

The results quoted in this paper all use the MRSA structure functions.
When using the CTEQ3M structure functions, the direct top cross section
increases by $15\%$ when the charm quark coupling is used,
corresponding to a $7\%$ improvement in the discovery limit for
$\kappa_c / \Lambda$.  This is primarily due to a larger charm quark
density in the proton with the CTEQ3M structure functions.
The $W + 1 \rm{~jet}$ cross section does not
change significantly, nor does the direct top cross section when the up
quark coupling is used.

We considered cases with and without b-tagging for each of the
possibilities in Table \ref{results}.  With the exception of Run 1 at
the Tevatron, b-tagging improved the discovery limit on $\kappa /
\Lambda$ by $2.0 - 2.5$ times.  However, for the data from Run 1 at the
Tevatron, b-tagging improves the discovery limit by only $40 \%$.  This
is mostly due to less efficient b-tagging, and to the smaller number of
events available with a lower luminosity.

In some single top quark production processes, there are regions of
overlap between, for example, $2 \rightarrow 1$ subprocesses and $ 2
\rightarrow 2$ subprocesses.  In particular, we worried about an overlap
between the direct top production and the gluon fusion diagram in which
one of the gluons is dissociated into a $c \overline{c}$ pair, and the
$c$ combines with the other gluon to produce a top quark.  Care must be
taken with these processes to avoid double counting.  A systematic
method exists for calculating a subtraction term which solves this
difficulty\cite{Heinson2,Barnett}.  The effect of the double counting is
most significant if the initial state particles are massive.  In the
case of direct top quark production due to anomalous t-c-g or t-u-g
couplings, the initial state particles are light enough that this does
not significantly affect the overall cross section.  We have therefore
ignored this effect in our calculation.

Although the background due to single top quark production (a top quark
with an associated jet) is small in the SM, there exists also the
possibility for single top quark production with the anomalous t-c-g (or
t-u-g) coupling, e.g. via $q \overline{q} \to t \overline{c}$ ($q
\overline{q} \to t \overline{u}$).  If the jet associated with the top
quark is not seen, this would enhance the direct top signal due to the
anomalous coupling.  Therefore, the discovery limits quoted in Table
\ref{results} are conservative estimates of the level to which $\kappa /
\Lambda$ may be probed.
A full treatment of single top production due to the anomalous t-c-g and
t-u-g couplings will be considered elsewhere.

In conclusion, we have calculated the discovery limits for the anomalous
chromomagnetic couplings t-c-g and t-u-g in hadron colliders using
direct production of an s-channel top quark.  We conservatively estimate
that an anomalous charm quark coupling can be detected down to $\kappa_c
/ \Lambda = .06 \rm{~TeV}^{-1}$ at Run 2 of the Tevatron, and 
$.009 \rm{~TeV}^{-1}$ at the LHC.  The cross section for the anomalous
up quark coupling is larger, and we can measure $\kappa_u / \Lambda$
down to $0.02 \rm{~TeV}^{-1}$ at Run 2 of the Tevatron, and $0.003
\rm{~TeV}^{-1}$ at the LHC.  The discovery limits for the upgraded
Tevatron are approximately two (six) times better than those obtained in
Ref. \cite{Han2} for $\kappa_c / \Lambda$ ($\kappa_u / \Lambda$).  The
relative size of the direct top production and the anomalous top decay
rate will help to differentiate the t-c-g and the t-u-g couplings.

Finally, we note that, in Ref \cite{Lopez}, the authors found that
electroweak-like corrections in a supersymmetric model can give Br($t
\to c g$) as large as $1 \times 10^{-5}$ for the most
favorable combinations of the parameters.  In terms of our
anomalous coupling parameter, this corresponds to $\kappa_c / \Lambda
= 0.0033$.  If supersymmetry is the only source for the anomalous t-c-g
coupling, our calculations therefore indicate that future improvements
at the LHC 
will be needed to make this a detectable signal, unless QCD-like
corrections \cite{Couture} further enhance the SUSY contributions, as
discussed in Ref. \cite{Lopez}.

\section*{Acknowledgments}
We would like to thank P. Baringer, J. Hauptman, A. Heinson, and X.
Zhang for many useful discussions.  This work was supported in part by
the U.S.~Department of Energy under Contract DE-FG02-94ER40817. M. Hosch
was also supported under a GAANN fellowship.


\begin{thebibliography}{99}
\bibitem{cdftop}F. Abe et al., Phys.Rev.Lett.74 (1995) 2626.
\bibitem{d0top}S. Abachi et al., Phys.Rev.Lett.74 (1995) 2632.
\bibitem{Incandela}J. Incandela, Nuovo Cim.109A (1996) 741.
\bibitem{LeCompte}Thomas J. LeCompte, FERMILAB-CONF-96-021-E (Jan 1996).
\bibitem{Heinson}A.P. Heinson, FUTURE TOP PHYSICS AT THE TEVATRON AND LHC.
    In *Les Arcs 1996, QCD and high energy hadronic interactions* 43-52.
\bibitem{Han1}T. Han, K. Whisnant, B.L. Young and X. Zhang, UCD-96-07
(Mar 1996), to be published in Phys. Rev. D; \\
K. Whisnant, Jin-Min Yang, Bing-Lin Young and X. Zhang,AMES-HET-97-1
(Feb 1997); \\
R. Martinez and J-Alexis Rodriguez,Phys.Rev.D55 (1997) 3212;
\bibitem{Rizzo}D. Atwood, A. Kagan and T.G. Rizzo, Phys.Rev.D52 (1995).
6264;\\ 
Douglas O. Carlson, Ehab Malkawi and C.P. Yuan, Phys.Lett.B337 (1994) 145;\\
G.J. Gounaris, M. Kuroda and F.M. Renard, Phys.Rev.D54 (1996) 6861;\\
G.J. Gounaris, J. Layssac, F.M. Renard, THES-TP-96-12 (Nov 1996).
\bibitem{Grzad}B. Grzadkowski,J.F. Gunion and P. Krawczyk, Phys. Lett.
B 268 (1991) 106; \\
G. Eliam, J.L. Hewett and A. Soni, Phys. Rev. D44 (1991) 1473;\\
M. Luke and M.J. Savage, Phys. Lett. B307 (1993) 387.
\bibitem{Couture}G. Couture, C. Hamzaoui and H. K\"{o}nig, Phys. Rev. D52
(1995) 1713.
\bibitem{Lopez}Jorge L. Lopez, D.V. Nanopoulos and Raghavan Rangarajan,
ACT-05-97 (Feb 1997).
\bibitem{Cheng}T.P. Cheng and M. Sher, Phys. Rev D48 (1987) 3484;\\
W.S. Hou, Phys. Lett. B296 (1992) 179;\\
L.J. Hall and S. Weinberg, Phys. Rev. D 48 (1993) 979;\\
D. Atwood and A. Soni, SLAC-PUB-95-6927.
\bibitem{Hill}C.T. Hill, Phys. Lett. B266 (1991) 419, B 345 (1995)
483;\\ 
B. Holdom, Phys Lett B339 (1994) 114, B351 (1995) 279.
\bibitem{Georgi}H. Georgi, L. Kaplan, D. Morin and A. Schenk, Phys.
Rev. D51 (1995) 3888;\\
For a review of composite models, see, e.g., R.D. Peccei in: Proc. 1987
Lake Louise Winter Institute: Selected Topics in the Electroweak
Interactions, eds. J.M. Cameron et al. (World Scientific, Singapore,
1987).
\bibitem{Zhang}X. Zhang, Phys. Rev. D51 (1995) 5309;\\
J. Berger, A. Blotz, H.-C. Kim and K Goeke, Phys. Rev. D54 (1996) 3598. 
\bibitem{Han2}T. Han, K. Whisnant. B.-L. Young and X. Zhang, Phys.
Lett. B385 (1996) 311.
\bibitem{MRS}A.D. Martin, R.G. Roberts and W.J. Stirling, Phys. Rev.
D50 (1994) 6734.
\bibitem{CTEQ} H.L. Lai, J. Botts, J. Huston, J.G. Morfin, J.F. Owens,
J. Qiu, W.K. Tung and H. Weerts;  Phys. Rev. D51, 4763(1995).
\bibitem{Heinson2}A.P. Heinson, A.S. Belyaev and E.E. Boos,
INP-MSU-96-41-448 (Dec 1996).
\bibitem{Barnett} R.M. Barnett, H.E. Haber and D.E. Soper, Nucl. Phys.
B306 (1988) 697; \\
F. W. Olness and W.-K. Tung, Nucl. Phys. B308, (1988) 813.
\bibitem{vecbos}F.A. Berends, H. Kuijf, B. Tausk and W.T. Giele, 
     Nucl.Phys. B357 (1991) 32. 
\end{thebibliography}
\end{document}